\documentstyle[12pt,aasms,tighten]{article}

\def\nofigure#1{\refstepcounter{figure}\label{#1}Figure~\ref{#1}.}
\def\icm{${\rm cm}^{-1}$}

\def\he3{$^3{\rm He}$}
\def\COBE{{\sl COBE}}

\def\IRAS{{\sl IRAS}}

\def\etal{{\it et~al.}}
\def\beq{\begin{equation}}
\def\eeq{\end{equation}}
\def\uK{\hbox{$\mu$K}}

\begin{document}
\sloppypar
\title{A Measurement of the Medium-Scale Anisotropy\\
in the Cosmic Microwave Background Radiation}

\author{
E.~S.~Cheng\altaffilmark{1},
D.~A.~Cottingham\altaffilmark{2},
D.~J.~Fixsen\altaffilmark{3},
C.~A.~Inman\altaffilmark{4},
M.~S.~Kowitt\altaffilmark{1},
S.~S.~Meyer\altaffilmark{4},
L.~A.~Page\altaffilmark{5},
J.~L.~Puchalla\altaffilmark{4},
and~R.~F.~Silverberg\altaffilmark{1}}
\altaffiltext{1}{NASA/Goddard Space Flight Center, Code 685.0, Greenbelt, MD 
20771}
\altaffiltext{2}{Global Science and Technology, Inc., NASA/GSFC Code 
685.0, Greenbelt, MD 20771}
\altaffiltext{3}{Applied Research Corporation, NASA/GSFC Code 685.3, 
Greenbelt, MD 20771}
\altaffiltext{4}{Massachusetts Institute of Technology, Room 20B-145, 
Cambridge, MA 02139}
\altaffiltext{5}{Princeton University Physics Dept., Princeton, NJ 08544}
\slugcomment{{\small Submitted to {\it Ap. J. Letters} 18 May 1993,
Revised 9 Nov 1993}}

\begin{abstract}
Observations from the first flight of the Medium Scale Anisotropy
Measurement (MSAM) are analyzed to place limits on Gaussian
fluctuations in the Cosmic Microwave Background Radiation (CMBR).
This instrument chops a 30\arcmin\ beam in a 3 position pattern
with a throw of $\pm40\arcmin$; the resulting data is analyzed in
statistically independent single and double difference datasets.  We
observe in four spectral channels at 5.6, 9.0, 16.5, and 22.5~\icm,
allowing the separation of interstellar dust emission from CMBR
fluctuations.  The dust component is correlated with the \IRAS\
100~\micron\ map.  The CMBR component has two regions where the
signature of an unresolved source is seen.  Rejecting these two
source regions, we obtain a detection of fluctuations which
match CMBR in our spectral bands of
$0.6 \times 10^{-5} < \Delta T/T < 2.2 \times 10^{-5}$ (90\% CL interval)
for total rms Gaussian
fluctuations with correlation angle 0\fdg5, using the single
difference demodulation.  For the double difference demodulation, the result is
$1.1 \times 10^{-5} < \Delta T/T < 3.1 \times 10^{-5}$ (90\% CL
interval) at a correlation angle of 0\fdg3.

\end{abstract}

\section{Introduction}
Measurements of the anisotropy of the Cosmic Microwave Background Radiation
(CMBR) on 0\fdg5 angular scales have received greater attention
(\cite{efstathiou92}, \cite{gorski93}, \cite{kashlinsky92})
after the \COBE\ detection on large angular scales (\cite{smoot92}).
The large-scale measurement sets an amplitude for the unperturbed primeval
density fluctuation spectrum;
at scales between 0\fdg5 and 2\deg, the anisotropy could be enhanced through
Doppler and heating effects.
The first flight of the Medium Scale Anisotropy Measurement (MSAM)
has led to a detection of brightness fluctuations at
0\fdg5 which are consistent with a CMBR spectral signature.
The instrument, observations, analysis, and results are described below.

\section{Instrument Description}
MSAM is a balloon-borne 30\deg\ off-axis Cassegrain telescope consisting
of a 1.4~m aluminum primary,
a 0.27~m nutating secondary, and a four channel radiometer previously
flown in
the Far Infra-Red Survey experiment (\cite {page89},
\cite{page90}, \cite{meyer91}).
The beam size of the telescope is 30\arcmin.
The secondary executes
a 2~Hz, four-position, square-wave chop ({\em i.e.} center, left, center, right)
with an
amplitude of $\pm40\arcmin$ on the sky.
Each position
consists of a 23~ms transition and a 102~ms stable integration where the
RMS position jitter is less than 4\arcsec.
The bolometric detectors have bandpass central frequencies of 5.6, 9.0, 16.5,
and 22.5 \icm\ each with $\sim 1$ \icm\ bandwidth (\cite{filt}).
The bolometer output is sampled
synchronously at 32~Hz with an integrating A/D converter.

The telescope is shielded
with aluminized panels
so that the dewar horn, the secondary and the geometrically
illuminated portion of the
primary have no view of the Earth
or the lower part of the flight train.
The edge of the balloon, as seen by the telescope, extends about
25\deg\ from the zenith.
Ground-based measurements of the sidelobe
pattern show the response to the ground to be less than $-60$~dB of the central
lobe of the antenna pattern.

\section{Observations}
The package was launched from Palestine, Texas at 0059~UT 5~June~1992, and
reached float altitude of 38~km at 0507~UT.
The flight ended with sunrise on the package at
1056~UT.  During the flight, we scanned Jupiter and Saturn to calibrate
the instrument and to map the antenna pattern, scanned over the center of the
Coma cluster to search for the Sunyaev-Zel'dovich effect (which will be
reported in a future {\sl Letter\/}), and integrated for 4.9~hours on a patch
of sky near the north celestial pole to search for CMBR anisotropy.

The CMBR observations are made by looking
8\deg\ above the north celestial pole,
and holding elevation constant while scanning in azimuth
$\pm 45\arcmin$ with a period of 1 minute.
We start with the center of the scan 21\arcmin\ to the east of the
meridian and track for
20~minutes; at this point the center of the scan is
21\arcmin\ to the west of the meridian.
We then pause for about 40~s to record an image with the star
camera and correct for gyro drift,
and begin a new 20~minute scan 42\arcmin\ to the east of the
previous scan.  Thus, half of each 20 minute
scan overlaps the following scan.
This observation strategy minimizes motion of the telescope relative to the
Earth and the atmosphere,
and therefore also minimizes any systematic contribution from these sources.
We completed 6 scans from 0419 to 0622~UT,
observed the Coma cluster for 40 minutes,
and completed an additional 8 scans
from 0701 to 0950~UT, for a total of 14 scans.  This covers two strips
at declination 82\fdg0, from right ascension 14\fh44 to 16\fh89, and
from 17\fh18 to 20\fh33 (J1992.5).  We refer to these two segments as
the two halves of the flight.

\section{Data Analysis}
\subsection{Calibration, Deglitching, and Demodulation}

The instrument is calibrated by in-flight observations of Jupiter before
the CMBR scans, and of Saturn after the CMBR scans.
We use Jupiter as our definitive calibration, with Saturn as a
comparison.  For the night of observation the apparent diameter of
Jupiter is 35\arcsec, and the diameter of the disk of Saturn is
17\farcs6.
We take the spectrum of Jupiter from \cite{Griffin86},
from which we derive antenna temperatures for our four channels
of 172, 170, 148, and 148~K respectively.
We assume that in our spectral region Saturn has a 
Rayleigh-Jeans spectrum with a temperature of
$133\pm20$~K (\cite {Harwit88}, \cite {Lang80}).
No correction is made for the rings of Saturn, which
at this frequency add less than 10\% in brightness (\cite{Hanel81}).
The calibrations on Jupiter and Saturn are consistent in channels 1,
3, and 4.  In channel 2, Saturn is dimmer than we expected by
a factor of two.  In addition, the offset in channel 2 drifts downward
by the same factor over the same period, and the noise also follows
this drift.  We assume that a post-detection gain drift is
responsible for this behavior, and we correct this by multiplying the
channel 2 data by a linear function of time to make the Jupiter and
Saturn calibrations consistent.
With this correction, the offset stability in channel 2 becomes consistent
with that of the other channels.
The uncertainty in the calibration is 10\% (\cite{Griffin86}).

The data contain large spikes, or glitches,
at a rate of once per 2 to 5~s, consistent with cosmic rays striking
the detectors (\cite{charak78}).
To remove these glitches, we perform a first
cut at $10\,\sigma$, and then deconvolve with a
model of the transfer function of the detectors.
The average offset signal in a chop cycle for each half of the flight
is then subtracted from all chop cycles in that half,
after which two cuts at $3.5\,\sigma$ are made.  About 5\%
of the data is cut this way.

We estimate instrument noise by forming the autocorrelation of the
deglitched data, filtering out 0~Hz and harmonics of 2~Hz, where sky
signals appear.  This estimate is therefore unaffected by any optical signal.
We form this estimate for each 20~minute segment of
data which is then propagated through the remaining
processing.  All $\chi^2$ reported below are with regard to these
error bars.

The data are demodulated in two different ways.  The first
corresponds to summing the periods when the secondary is in the central
position, and subtracting the periods when it is to either side; we
call this the double-difference demodulation.  This
demodulation is insensitive to atmospheric gradients.
The second demodulation corresponds to differencing the periods when the
secondary is to the right from those when it is to the left, and
ignoring the periods when the secondary is in the center; we call this
the single-difference demodulation.
These two demodulations of the data yield
statistically independent measurements of the sky.
We use the scan over Jupiter to deduce an optimal demodulation of the
infrared signal.
Each group of four complete chopper cycles (2~s) is averaged together
to form a ``record.''
Records for which there are too few samples to form a robust average,
due to deglitching or telemetry dropouts, are deleted.
The procedure results in a 2\% data loss.
Each record is then demodulated to produce one single and one double
difference observation every 2~s.

This demodulated signal has had a constant offset removed from it in
the deglitching process.  The size of this offset is approximately
10~mK Rayleigh-Jeans in all four bands, and in both the single and
double difference demodulation.
To remove the drift in this offset, we subtract a slowly varying
function of time to minimize the variance of observations of each point
in the sky made at different times.
The function of time is implemented as a cubic
spline with a knot every 2.5 minutes.
Each channel and demodulation is treated separately.
The two halves of the flight are also treated
separately, as they are separated in time and do not overlap on the
sky.
The drift is $\la 400\uK$ Rayleigh-Jeans, and
significant at the 3--8$\,\sigma$ level;
thus the offset is constant to about 4\%.
The reduced $\chi^2$ of these fits range from 0.92 to 1.21.

The data are binned by position on the sky, and by relative angular
orientation of the antenna beam with respect to the local tangent
to the circle of constant declination at the central beam location.
The bin size is 0\fdg12 in position, and
10\deg\ in angular orientation.
Records which differ from the
median in the bin by more than $3\,\sigma$ are deleted.  Following this,
bins containing fewer than 10
records are deleted.
The reduced $\chi^2$ of the binned data after removing a mean
from each bin
ranges from 0.88 to 1.04 for the various
channels and demodulations,
indicating that our observations of the sky are consistent.
The binned data contains 86\% of all the data originally taken,
with an achieved sensitivity in each of the four channels of
400, 210, 140, and 330~\uK~$\sqrt{\rm s}$ Rayleigh-Jeans.
For channels 1 and 2 this is 810 and 1190~\uK~$\sqrt{\rm s}$ CMBR.

\subsection{Spectral Model of the Sky}
To extract the part of the signal due to variations
in the CMBR, we fit the data $t_{ck}$
for each channel $c$ and sky bin $k$ to
a two component model:
\beq
t_{ck} = \int \, d\nu\, F_c(\nu)\left[
D_k \left(\frac{\nu}{\nu_0}\right)^\alpha B_\nu(T_{\rm D})
+ t_k \left. \frac{dB_\nu}{dT}\right|_{T_{\rm CMBR}} \right],
\label{e_decomp}
\eeq
where $F_c(\nu)$ is the spectral response of the instrument,
$B_\nu(T)$ is the Planck function at temperature $T$,
$T_{\rm D} = 20 \, {\rm K}$ is the dust temperature,
$\alpha = 1.5$ is the spectral index of the dust,
$\nu_0 = 22.5$ \icm\ is the reference frequency,
$T_{\rm CMBR} = 2.73 \, {\rm K}$ is the temperature of the CMBR,
and $D_k$ and $t_k$ are free parameters.
The result is a component sensitive to the CMBR ($t_k$)
and a component sensitive to the dust ($D_k$).
The $\chi^2$ for the fit is $237/292$ for the single difference data,
and $454/294$ for the double differenced data.
$T_{\rm D}$ and $\alpha$ are fixed for this analysis since varying these
parameters by reasonable amounts does not significantly change the CMBR
component.
Fig.~\ref{f_dust} shows the dust channel.
The superimposed curve is an approximation of the expected dust emission
produced by convolving our antenna pattern with IRAS
100~\micron\ measurements (\cite{IRAS91}, \cite{IRAS93}).
For the most part agreement is quite good, but in a few places
they differ quite significantly.
The scaling of the \IRAS\ data is determined by fitting to the dust channel;
this scaling is equivalent to an average
spectral index between 100~\micron\ and 444~\micron\ (22.5~\icm) of
$1.5\pm0.2$.
Fig.~\ref{f_cmbr} shows the CMBR component.
For clarity, these plots are binned more coarsely than
the data we analyzed, and do not distinguish between points taken at
slightly different declination or antenna orientation.

There are two candidate unresolved sources visible in the CMBR spectral
component, $\Delta T_k$, for both the single and double difference data.
The more prominent source (MSAM 15+82) is located at
RA 14\fh92~$\pm$~0\fh03 in a dust-free region, and has a
measured flux density of $4.5\pm0.7$ Jy at 5.6~\icm\ (error bar
includes calibration uncertainty of 10\%).
The second, dimmer, candidate (MSAM 19+82) is located at
RA $19\fh29 \pm 0\fh03$, and has a measured flux density of
$3.6 \pm 0.6$~Jy at 5.6 \icm.
It is in a region that is somewhat confused by foreground dust emission.
We observe at a fixed declination so the declination coordinate for these
sources is less well determined,
but it is most likely within a beamwidth of the declination of observation,
82\fdg00~$\pm$~0\fdg25.
Each of these sources has a signal which is stationary with respect to the
sky (as determined by the various levels of chopping built into our
observation strategy) and is detected in multiple channels.
The compactness of these sources makes it implausible that they are
due to diffraction or side-lobe effects.

We cannot rule out the possibility that these are are Galactic
bremsstrahlung sources;
observations at lower frequency ($\la 3$~\icm) will shed light on this
question.
It would be somewhat unexpected for CMBR fluctuations obeying Gaussian
statistics to produce such features, and we have performed simulations that
indicate that these features are not consistent with the correlation functions
considered here.
For the detailed discussion in this paper,
we have made the assumption that these are indeed unresolved foreground
sources.
The regions which are contaminated by these sources have been removed from
consideration pending further analysis.
In particular, only the region 15\fh69~$<$~RA~$<$~18\fh55 is included
in the main CMBR results, though we also include results based the
entire data set.
A future {\sl Letter} will address the detailed spectra and possible
identification of these sources.

\subsection{Limits on CMBR Anisotropy}
\def\x{{\bf x}}
To set limits on the anisotropy of the CMBR,
we model the anisotropy $\Delta T(\x)$ as a
Gaussian random field described by a correlation function $C(|\x_1-\x_2|) =
\langle \Delta T(\x_1)\,\Delta T(\x_2) \rangle$.  Our observations
have been binned by orientation of the antenna pattern; call the
antenna pattern as oriented for the $k$th observation $H_k(\x)$.
(Clearly all the
functions $H_k$ are translations and rotations of one function $H$.)
Then the signal $s_k$ from CMBR is
$s_k = \int d\x\,H_k(\x)\Delta T(\x),$
and consequently the covariance of the $s_k$ is
\beq
\langle s_k s_l \rangle = \int
d\x_1\,d\x_2\,H_k(\x_1)H_l(\x_2)C(|\x_1-\x_2|).
\eeq
To this signal our instrument adds noise $n_k$, which has the
covariance $\langle n_k n_l \rangle = \delta_{kl}\sigma^2_k$.  The
instrument noise and sky signal are uncorrelated with each other, so
the covariance of our observations $t_k \equiv s_k + n_k$ is just the
sum of the covariances of $s_k$ and $n_k$.

We set limits on the overall amplitude of the correlation function
$C$ by using the likelihood ratio statistic (\cite{Martin71}).
Let $(W_{kl})^{-1} = \langle s_k s_l \rangle + \sigma^2_k\delta_{kl}$,
and $(W^0_{kl})^{-1} = \sigma^2_k\delta_{kl}$;
then the likelihood ratio
$\lambda$ is
\beq
\lambda = \left(\frac{\det W}{\det W^0}\right)^{1/2}
\exp\left(-{1\over2} \sum_{kl} t_k (W_{kl} - W^0_{kl}) t_l \right).
\eeq
Let $\rho_C(\lambda)$ be the probability density function of $\lambda$ under
the hypothesis that the fluctuations obey the correlation
function $C$,
and let $\lambda^*$ be the value of the statistic
for our observations.
Note that $\lambda$, $\lambda^*$, and $\rho_C$ all depend on the
correlation function, and in particular on its amplitude.
Then the 95\% confidence level upper limit on the
amplitude is that amplitude for which the cumulative distribution function is
\beq
\int_0^{\lambda^*} \,d\lambda\,\rho_C(\lambda) = 0.95.
\label{e_lambda_int}
\eeq
Similarly, the 95\% lower limit is that amplitude for which this
integral is 0.05, if such an amplitude exists.  Taken together, these
two limits form a 90\% confidence interval.
We perform the integral in (\ref{e_lambda_int}) by Monte-Carlo
integration (Press \etal\ 1986).
The amplitudes we quote here are total rms fluctuation, i.e.,
$[C(0)]^{1/2}$.

Fig.~\ref{f_dt} shows the upper and lower limits for total rms anisotropy
as a function of correlation angle and
assuming a Gaussian-shaped correlation function.
This uses only the data in the region 15\fh69~$<$~RA~$<$~18\fh55.
The single difference data is most sensitive at
$\theta_c = 0\fdg5$ with a 90\% 
confidence interval of 
$0.6 \times 10^{-5} < \Delta T/T < 2.2 \times 10^{-5}$.
For the double difference data, the most sensitive result is at
$\theta_c = 0\fdg3$,
where $1.1 \times 10^{-5} < \Delta T/T < 3.1 \times 10^{-5}$.
We have analyzed various subsets of the data, dividing the flight into
unequal quarters.  The first quarter is the region near MSAM 15+82,
the second the following data up to the point where we moved off to
observe the Coma cluster, the third the source-free data after the
break, and the fourth the data near MSAM 19+82.
Table~\ref{t_dt} gives the upper and lower limits
for each quarter of the flight
as well quarters 2 and 3 on which these results are based.
We emphasize that our observation strategy allows for independent
measurements using the single and double difference demodulations.

\section{Conclusions}

The results from both the single and double difference data for this flight
(declination 82\deg\ and 15\fh69~$<$~RA~$<$~18\fh55)
show positive detections of sky brightness variations which
are consistent with a CMBR spectrum.
The placement of our spectral passbands allows for strong discrimination
of warm Galactic dust from CMBR fluctuations, but some cold dust models
at low levels are difficult to rule out.
We cannot strongly rule out the spectrum of bremsstrahlung,
though it is highly unlikely that that could cause a signal of the measured
amplitude.
\cite{Meinhold93} have previously reported an upper limit
($\Delta T/T < 2.5\times10^{-5}$) at this angular scale.

The presence of the two unresolved sources in the data
is clearly very interesting, and could imply a significantly different
scenario for CMBR anisotropy measurements on 0\fdg5 angular scales
at the $\Delta T/T \sim 1\times10^{-5}$ level.
If the extrapolations of \cite{franceschini89} are correct, then
it is not likely that this detection is due to unresolved extragalactic
sources.
However, the two sources in these data may indicate that
there is a previously unsuspected population for which neither the
spatial distribution nor the spectral signature is well determined
by existing data.
Alternatively, if these sources are true CMBR fluctuations, then we need
to investigate the compatibility of these highly peaked features with
various models.
Clearly, a more complete understanding of these sources is central to
further improvements in sensitivity to CMBR anisotropies.

Files containing the data on which Fig.~\ref{f_cmbr} is based
along with the antenna pattern will be made available.
For information on how to obtain these data, fetch the file
{\tt /pub/data/msam-jun92/README} by anonymous FTP from
{\tt cobi.gsfc.nasa.gov}.

\acknowledgments
We would like to thank the staff of the National Scientific Balloon
Facility in Palestine, Texas, who valiantly maintain the original
can-do spirit of NASA.  W.~Folz provided electronics support and
accompanied us on our observing run in Palestine.
We would like to
thank C.~Lisse, J.~Mather, H.~Moseley, and R.~Weiss for helpful discussions.
The NSF Research and NASA Science Internet,
which we take for granted too often,
makes our long-distance collaboration possible without exorbitant expenditure
of travel funds.
This work made extensive use of the fine products of
the Free Software Foundation of Cambridge, MA.
The National Aeronautics and Space Administration
supports this research through grants  NAGW 1841, RTOP 188-44, NGT 50908, and
NGT 50720.
Lyman Page is supported by a grant from the National Science Foundation.

\clearpage
\begin{table}
\begin{center}
\begin{tabular}{lccccc}
&$\theta_c$&Quarter&RA&Upper bound&Lower bound\\
&&&(h)&(\uK)&(\uK)\\
\tableline
Single&0\fdg5&1&14.44--15.69&221&39\\
difference&&2&15.69--16.89&   99&---\\
&&3&17.18--18.55&            102&15\\
&&4&18.55--20.33&            221&62\\
&&2+3&15.69--18.55&           61&16\\
&&All&14.44--20.33&          116&53\\
Double&0\fdg3&1&14.44--15.69&336&91\\
difference&&2&15.69--16.89&  139&33\\
&&3&17.18--18.55&            127&21\\
&&4&18.55--20.33&            121&37\\
&&2+3&15.69--18.55&           85&30\\
&&All&14.44--20.33&           97&50\\
\end{tabular}
\end{center}
\caption{Upper and lower bounds on CMBR anisotropy}
\label{t_dt}
\end{table}

\clearpage
\nofigure{f_dust} Optical depth at 22.5~\icm\ of
dust component $D_k$;
a) Double difference demodulation, b) Single difference.
Superimposed curve is \IRAS\ 
100~\micron\ data convolved with our antenna pattern;
scale is set by fit to our observations.

\nofigure{f_cmbr} CMBR component $t_k$; a) Double difference, b) Single
difference.
Antenna pattern is superimposed for reference.

\nofigure{f_dt} Upper and lower limits on CMBR anisotropy for Gaussian-shaped
power spectra with correlation length $\theta_c$, based on data
between right ascensions 15\fh69--18\fh55.
95\% CL upper limit for double difference (solid) and single difference
(long dashed);
95\% CL lower limit for double difference (dashed) and single difference
(dotted).

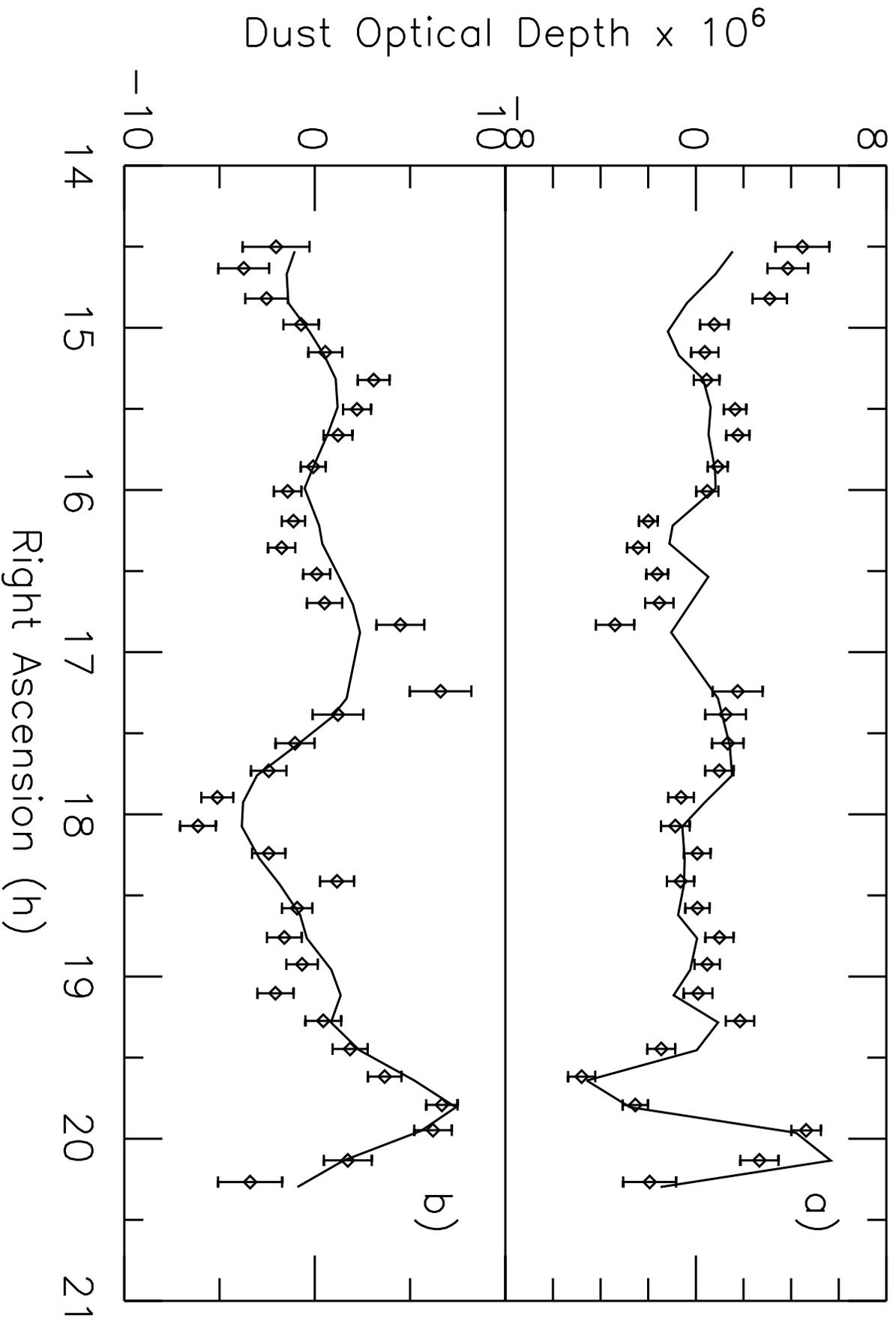
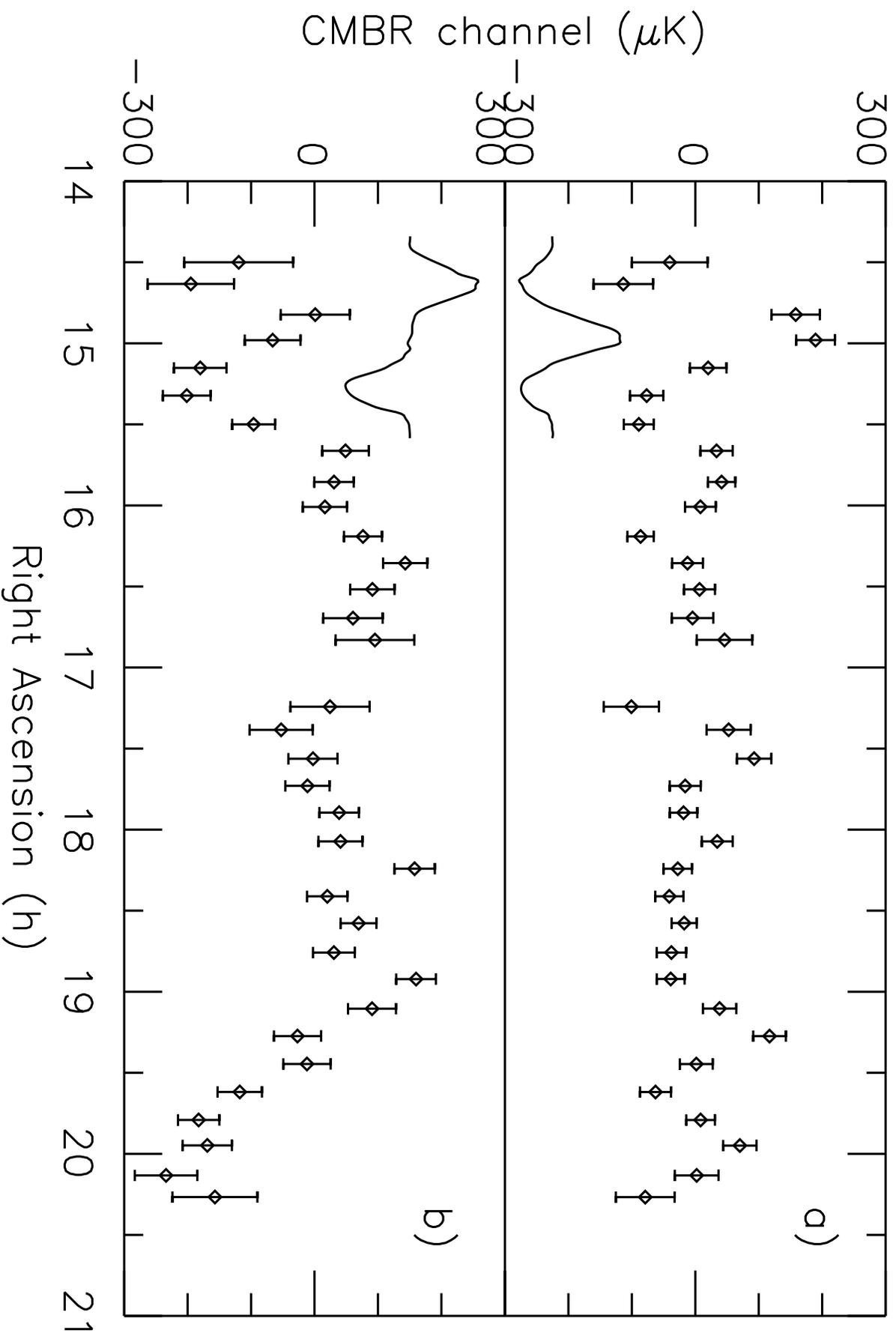
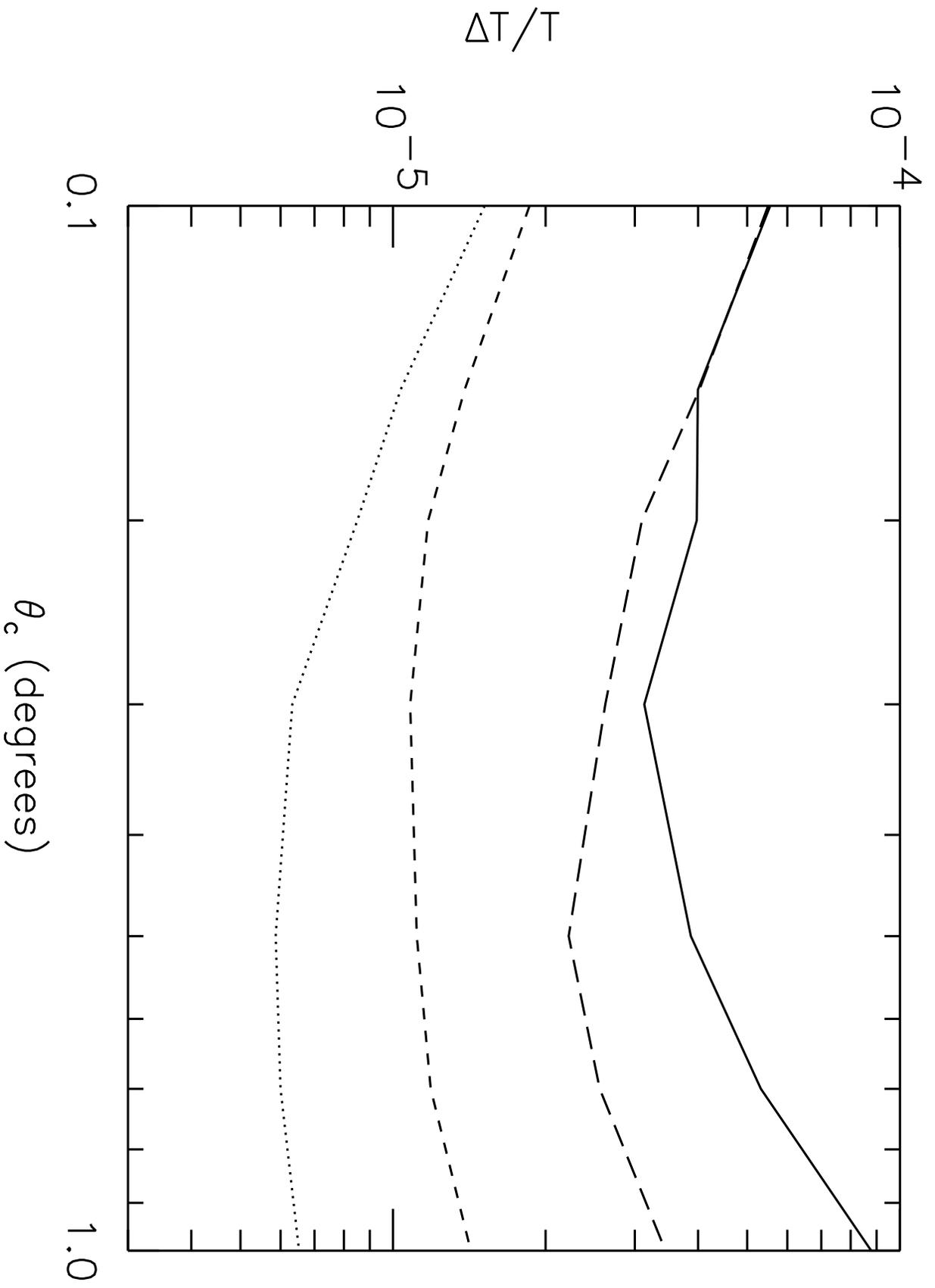
\end{document}